%% file: main.tex
\documentclass[conference,9pt]{IEEEtran}
\usepackage{lipsum}
\usepackage{float}
\usepackage{footmisc}
\usepackage{cite}
\usepackage{amsmath,amssymb,amsfonts}
\usepackage{algorithmic}
\usepackage{color}
\usepackage{graphicx}
\usepackage{subfigure}
\usepackage{textcomp}
\usepackage{theorem}
\usepackage{multirow,booktabs}
\usepackage[linesnumbered,ruled,vlined]{algorithm2e}
\usepackage{amsmath} 
\UseRawInputEncoding
\usepackage{etoolbox}
\usepackage[T1]{fontenc}
\usepackage[utf8]{inputenc}
\usepackage[normalem]{ulem}

\allowdisplaybreaks

\makeatletter
\patchcmd{\@makecaption}
  {\scshape}
  {}
  {}
  {}
\makeatletter
\patchcmd{\@makecaption}
  {\\}
  {.\ }
  {}
  {}
\makeatother

\newcommand{\qdel}[1]{}
\newcommand{\xadd}[1]{{\color{black}{#1}}}
\newcommand{\xdel}[1]{}
\SetKwInput{KwData}{Function}

\newtheorem{Example}{Example}

\title{\huge High-Quality Iterative Logic Compiler for In-Memory SIMD Computation with Tight Coupling of Synthesis and Scheduling}
\author{
    \IEEEauthorblockN{Xingyue Qian$^{1}$, Chenyang Lv$^{2}$, Zhezhi He$^{2,*}$, and Weikang Qian$^{1,3,*}$}
    \IEEEauthorblockA{$^{1}$University of Michigan-SJTU Joint Institute, $^2$School of Electronic Information and Electrical Engineering, \\and $^3$MoE Key Lab of AI, Shanghai Jiao Tong University, Shanghai, China}
    \IEEEauthorblockA{Emails: \{qianxingyue, lvchenyang, zhezhi.he, qianwk\}@sjtu.edu.cn; $^*$corresponding authors}
}

\IEEEoverridecommandlockouts
\begin{document}
\maketitle

\begin{abstract}
\input{0_abstract}
\end{abstract}

\input{1_introduction}

\input{2_preliminaries}

%\input{3_related}

\input{4_target}

\input{5_method}

\input{6_results}

\input{7_conclusion}
% \newpage
\bibliographystyle{IEEEtran}
\bibliography{cite}
\end{document}

%% file: 0_abstract.tex
In-memory computing (IMC) with single instruction multiple data (SIMD) setup enables memory to perform operations on the stored data in parallel to achieve high throughput and energy saving.
To instruct a SIMD IMC hardware to compute a function, a logic compiler is needed that involves two steps: logic synthesis and scheduling.
Logic synthesis transforms the function into a netlist of supported operations.
Scheduling determines the execution sequence and memory location of the operations and outputs the instruction sequence given to the hardware.
In this work, we propose an iterative logic compiler with tight coupling of synthesis and scheduling to find high-quality instruction sequences.
It is based on improving the critical sub-netlist identified by our algorithm and performing problem-specific resubstitution.
The experimental results show that our compiler can obtain better instruction sequences with energy-delay products reduced by 18.0\% on average compared to the best state-of-the-art method.

%% file: 1_introduction.tex
\section{Introduction}\label{sec:intro}
The data transfer between processor and memory in conventional von Neumann architecture is a performance bottleneck~\cite{WALL}.
To reduce data transfer,
\emph{in-memory computing (IMC)} is proposed, which enables memory to perform primitive logic operations such as majority (MAJ) and XOR on the stored data~\cite{MAGIC,SIMDRAM,PLIM,XMG-GPPIC}.
\emph{Single instruction multiple data (SIMD)} is a popular design style of IMC with high throughput~\cite{SIMDRAM,28NM,PLIM,XMG-GPPIC}, where all computation of a function, \textit{e.g.}, addition, on an input pattern is performed in a single column of the memory array.
In this way, by processing the cells in different columns of the same row \emph{bitwise}, different columns can be used to compute the same function with different input patterns in parallel.
\begin{figure}[!htbp]
\centering
\includegraphics[scale=0.34]{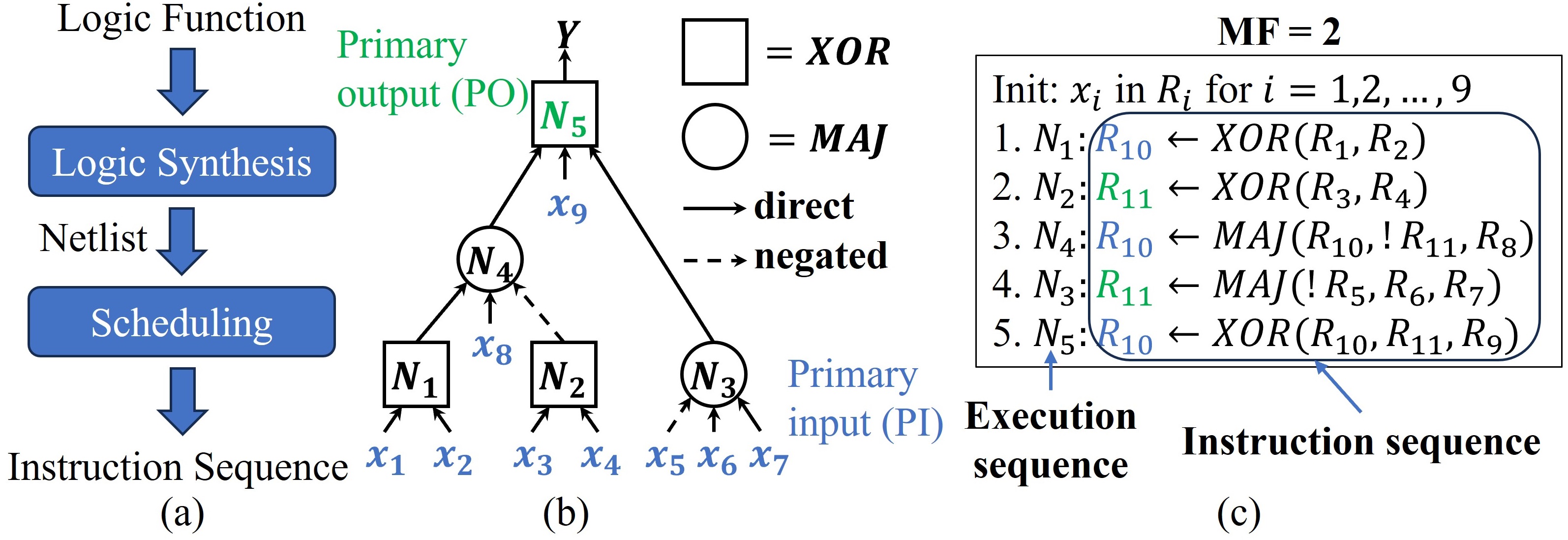}
\caption{(a) The flow of a SIMD IMC logic compiler; (b) a netlist example. Note that some operations in the figure only have two fan-ins, because we omit the constant 0 for simplicity; (c) a scheduling result example.}
\label{fig:BC}
\end{figure}

To instruct a SIMD IMC hardware to compute a function on the stored data, a logic compiler is needed to obtain the \emph{instruction sequence} given to the hardware~\cite{XMG-GPPIC,MIG,SIMPLER,STAR,PREV,LOGIC,SYN}.
An instruction is either performing a supported operation on some rows within a memory array or doing some data management such as copying one row to another.
We call an instruction sequence that instructs the hardware to compute a function a \emph{design} of the function.
Different designs have different energy-delay products (EDPs).
The task of a logic compiler is to find a design with a low EDP.
The basic logic compiler flow is shown in Fig.~\ref{fig:BC}(a). It has two steps: \emph{logic synthesis} and \emph{scheduling}.
Logic synthesis converts the target function into a netlist of the supported operations, \textit{e.g.}, XOR-majority graph (XMG) as shown in Fig.~\ref{fig:BC}(b).
Since it takes time and energy to perform each operation in the netlist in serial in the SIMD setup, a sub-target in the logic synthesis step is to minimize the \emph{size} of the netlist, \textit{i.e.}, the number of operations.
Then, the scheduling step determines the \emph{execution sequence} of the operations and the rows to store their results, and outputs the instruction sequence as shown in Fig.~\ref{fig:BC}(c).
The number of rows required to store the results of the operations is called \emph{memory footprint (MF)}.
A common sub-target of the scheduling step is to minimize MF, since designs with less MF allow more operations to be performed within a single memory array, reducing the amount of costly cross-array data copying~\cite{XMG-GPPIC,SIMPLER,CHAL}.

Most existing logic compilers are based on this basic flow~\cite{MIG,SIMPLER,STAR,LOGIC,SYN,PREV}.
In the logic synthesis step, the function is first converted into a supported netlist, and then a sequence of logic optimization commands is applied to the netlist to reduce its size.
Most works use existing commands such as rewriting~\cite{REW1}, which are directly taken from logic synthesis tools such as ABC~\cite{ABC}.
A few other works modify the existing commands to meet the specific requirement of their target hardware~\cite{MIG}.
% For example, in~\cite{MIG}, the authors modify the majority-inverter graph (MIG) optimization algorithm so that the number of operations with multiple complemented inputs is reduced, since performing those operations requires extra clock cycles and rows on the target hardware~\cite{PLIM}.
In the scheduling step, the optimized netlist is converted to an instruction sequence, using a heuristic~\cite{MIG,SIMPLER,STAR} or a divide-and-conquer~\cite{PREV} approach to minimize MF.
These basic compilers~\cite{MIG,SIMPLER,STAR,LOGIC,SYN,PREV} are relatively fast, but they can only produce a single design with sub-optimal performance. Besides, they make no attempt to improve the performance of the design.
Recently, reinforcement learning (RL) is used to obtain a sequence of optimization commands that is applied to a baseline netlist to improve its EDP~\cite{XMG-GPPIC}.
However, due to the long runtime of the RL algorithm, the sequence only contains 10 commands, and the scheduler used in the process is a simple one.
Hence, the design obtained is still unsatisfactory.

To address the issues of the existing logic compilers, we propose an iterative logic compiler with a tight coupling of logic synthesis and scheduling. Our contributions are:
\begin{itemize}
\item For the first time, we reveal that a netlist with a smaller size does not necessarily has less MF and less EDP.
Based on this fact, we propose to find Pareto-optimal designs in terms of the two metrics, \textit{i.e.}, size and MF.
\item We propose a flow that iteratively picks a Pareto-optimal design obtained so far and attempts to improve its performance by re-synthesizing and re-scheduling it.
\item
To speed up the iterative flow, we propose a technique exploiting the scheduling result to extract a critical sub-netlist for re-synthesis instead of the whole netlist.
\item To further improve the performance of the compiler, we propose a powerful logic optimization method based on traditional resubstitution~\cite{SIMRESUB}, which directly targets at reducing MF by exploiting the scheduling result.
\item We also make the code of our compiler open-source at 
https://github.com/SJTU-ECTL/IMCCompiler.
% https://anonymous.4open.science/r/IMCCompiler-ED42.
\end{itemize}

The experimental results show that our compiler can obtain designs with EDP reduced by 18.0\% on average compared to the best state-of-the-art method.

% The experimental results show that our compiler can obtain better Pareto-optimal designs on size and MF, and the EDPs of the designs are reduced by 18.0\% on average compared to the best state-of-the-art method.

% The rest of the paper is organized as follows.
% Section~\ref{sec:background} provides the background.
% Section~\ref{sec:probform} presents our study on the relationship among size, MF, and EDP.
% Section~\ref{sec:method} presents the proposed iterative compiler featuring tight coupling of synthesis and scheduling.
% Section~\ref{sec:result} shows the experimental results.
% Section~\ref{sec:conclusion} concludes the paper.

%% file: 2_preliminaries.tex
\section{Preliminaries}\label{sec:pre}
% In this section, we introduce some preliminaries.
\subsection{Netlist Representation and Basic Scheduling Rules}\label{sec:pre_comp}
In this work, we use the architecture proposed in~\cite{XMG-GPPIC} as our target architecture.
It supports inversion and 3-input XOR and MAJ operations, where the inversion is achieved by directly reading the negated output of a memory cell.
Hence, the corresponding netlist is an XMG as shown in Fig.~\ref{fig:BC}(b).
Each node in the netlist corresponds to a supported operation, \textit{i.e.}, either XOR or MAJ, and we use the terms operation and node interchangeably.
A directed edge from node $i$ to node $j$ indicates that the output of node $i$ is an input of node $j$.
If the edge is dashed, the output of node $i$ is negated.
We call node $i$ a \emph{fan-in} of node $j$ and node $j$ a \emph{fan-out} of node $i$.
The \emph{primary inputs (PIs)} of the function, \textit{e.g.}, the blue variables in Fig.~\ref{fig:BC}(b), are initially stored in memory and cannot be deleted during the process, since they may still be needed in the future.
For the \emph{primary outputs (POs)} of the function, \textit{e.g.}, node $N_5$ in Fig.~\ref{fig:BC}(b), their results must be in memory when the computation ends, and thus also cannot be deleted once computed.
The results of other operations can be deleted from the memory once they are no longer needed to save memory space.
Here, by deleting an operation result from memory, we mean that the row containing the result can be overwritten and hence, becomes available for future operations.
% Note that some operations in the figure only have two fan-ins, because we omit the constant 0 for simplicity.

An operation can be executed by passing an instruction to the hardware when all its fan-ins are stored in memory.
In the example in Fig.~\ref{fig:BC}(c), the execution sequence is $N_1, N_2, N_4, N_3, N_5$, and the corresponding instruction to execute each operation is listed to the right of `:'.
For example, the instruction at clock cycle three is $R_{10}\leftarrow \textit{MAJ}(R_{10},!R_{11},R_8)$ to perform operation $N_4$, meaning that we perform an MAJ operation on row 10, negated row 11, and row 8, which contain the results of operation $N_1$, operation $N_2$, and PI $x_8$, respectively, and store the result in row 10.

However, an operation can only be applied to the rows in the same memory array, but the number of rows in an array is limited~\cite{NVSIM}.
When the rows containing the fan-ins of an operation are from different arrays, we need to execute extra costly cross-array data copying instructions before executing the operation.
This may diminish the advantage of IMC~\cite{XMG-GPPIC,SIMPLER,CHAL}.
Hence, the common target of scheduler is to minimize the MF~\cite{MIG,XMG-GPPIC,SIMPLER,STAR,LOGIC,PREV} so that we can compute functions with larger netlists within a single array to reduce EDP. Also, when multiple arrays are required, \xdel{since }the designs with less MF tend to require less copy instructions, thus having less EDP.
% Our experiment with several large netlists in Fig.~\ref{fig:form}(a) confirms this point, where we show the EDPs of the same netlist of each function scheduled by different schedulers.
% The EDP is evaluated using the method in~\cite{XMG-GPPIC}, accumulating the energy and delay of each instruction, and the number of rows in an array is 256.
% For the same netlist, designs with less MF tend to have less EDP.

Given an execution sequence of the netlist, we can easily obtain the corresponding minimum MF by \xdel{deleting the operation results from the memory once they are not needed \xdel{any more }and }storing the result of the scheduled operation at each clock cycle in the available row with the smallest index. Note that a row becomes available once the operation result stored in it is not needed any more. Hence, the main target of the scheduler is to find an execution sequence with less corresponding MF.

\subsection{Resubstitution}\label{sec:back_resub}\label{sec:pre_resub}
Resubstitution is a powerful method in logic synthesis for reducing the size of a netlist~\cite{SIMRESUB}.
Its main idea is to check whether the function of a node, called \emph{root}, in the netlist can be expressed using other existing nodes, called \emph{divisors}.
If so, we substitute the root with the expression we find, and all nodes that are used exclusively by the root, \textit{i.e.}, its \emph{maximum fanout free cone (MFFC)}~\cite{WINDRESUB}, can be deleted, thus reducing the size of the netlist.
A state-of-the-art resubstitution method is the simulation-guided resubstitution~\cite{SIMRESUB}.
% It first simulates the netlist with some \qdel{selected }input patterns and stores the simulation results.
It traverses the netlist, treating each node as the root, and collects a subset of the nodes in the netlists as its potential divisors.
It checks for the existence of an expression based on logic simulation followed by validation via a satisfiability solver.
% The expression is usually either a divisor itself or a supported operation with the divisors as its inputs.
% For example, in XMG, we check whether $r=d_1$, $r=\textit{XOR}(d_1,d_2,d_3)$, and $r=\textit{MAJ}(d_1,d_2,d_3)$, where $r$ is the root\qdel{ node}, and $d_1,d_2,d_3$ are the possibly complemented divisors.
% When such an expression exists, its function still may differ from that of the root node since there are input patterns not used in the simulation.
% We call the expression found in this way a \emph{candidate}, and a satisfiability (SAT) validator is used to check whether a candidate is really functionally equivalent to the root\qdel{ node}.
% The candidate is used to substitute the root \qdel{node }only if it passes the equivalence checking.
% The process can be further improved by using don't cares and a normalization technique~\cite{SIMRESUB}.

%% file: 4_target.tex
\section{Motivation}\label{sec:mot}

\begin{figure}
\centering
\includegraphics[scale=0.45]{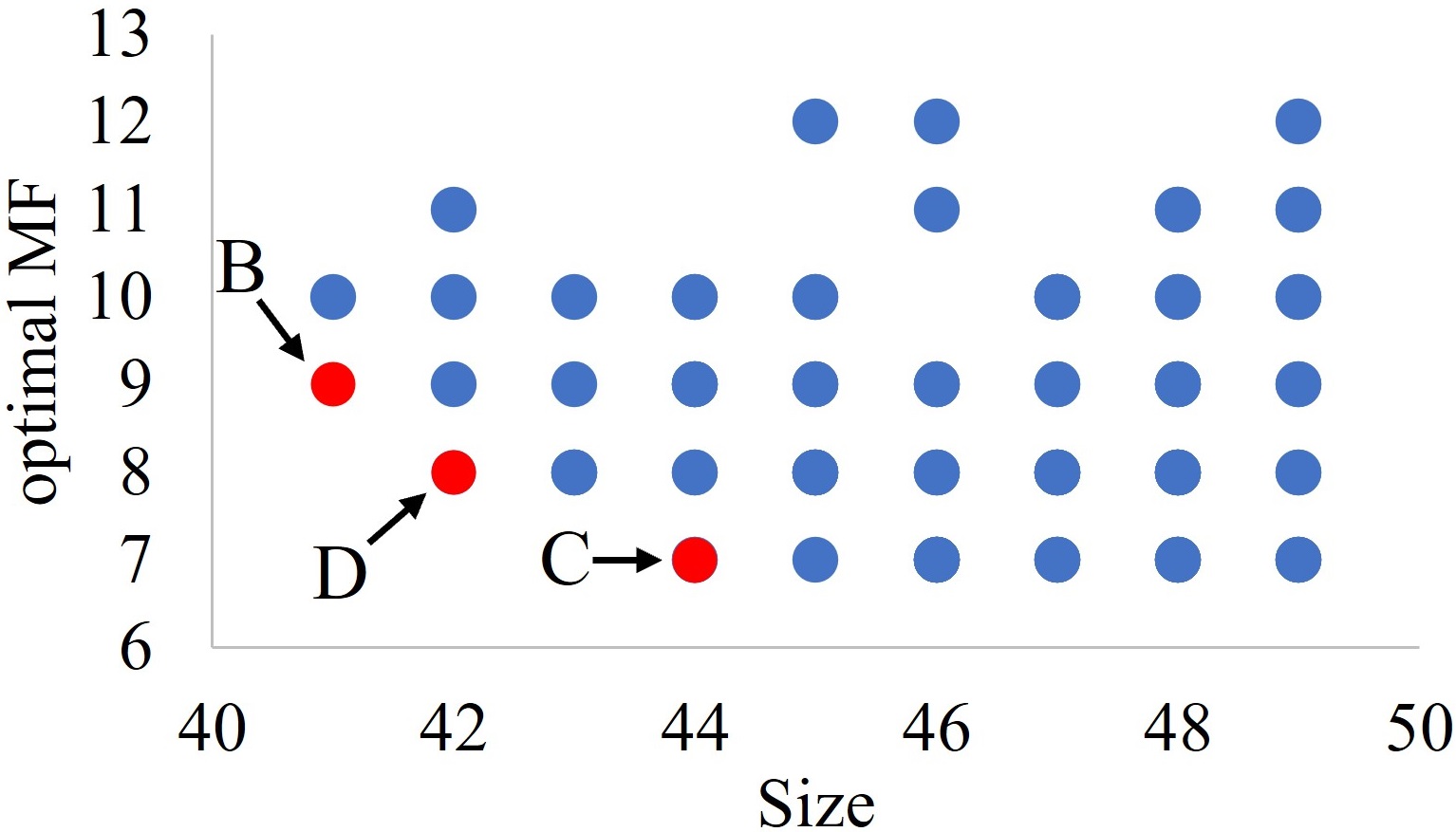}
\caption{\xdel{(a) The relation between $\textit{EDP}$ and $\textit{MF}$. (b) d}Design space of \textit{x2} function on MF-size plane.}
\label{fig:form}
\end{figure}

As we mentioned, designs with less size and less MF tend to have less EDP after implemented on hardware.
Hence, the basic logic compiler flow first performs synthesis, targeting at minimizing the netlist size, and then performs scheduling, targeting at minimizing the MF.
However, in our experiments, we found that netlists with smaller sizes do not necessarily have less MF, and thus \xdel{does}\xadd{do} not necessarily have less resulting EDP.
This means that the design obtained by the basic compiler flow does not necessarily have the least EDP.
\vspace{-1em}
\begin{Example}
In the example in Fig.~\ref{fig:form}, \xadd{each point corresponds to a design of the \textit{x2} function. }\xdel{the}\xadd{The} best design found by the basic flow\xadd{, \textit{i.e.}, first minimizing size then minimizing MF,} would be design B. It has 41 operations and\xadd{ an} MF of 9.
If we use arrays with 8 available rows for the computation, it requires two arrays and at least one copy instruction.
In contrast, although design D has one more operation, its MF is only 8, meaning that it only requires one array and does not need any copy instructions.
Although a copy instruction takes one cycle like an operation instruction, it consumes much more energy than an operation\xadd{ instruction}~\cite{XMG-GPPIC}, so design D will have less EDP than design B found by basic compiler.
\end{Example}

To deal with the non-positive correlation between size and MF, instead of minimizing the two metrics in serial in synthesis and scheduling steps, respectively, we propose to find the Pareto-optimal designs on size and MF of the given function.
% Note that a design $A$ is \emph{Pareto-optimal} if there does not exist another design $B$ such that its size and MF are smaller than or equal to those of $A$, respectively, and at least one metric of $B$ is strictly smaller than that of $A$.
%
Once we obtain the Pareto-optimal designs in terms of size and MF, we can quickly choose the one with the least EDP among them based on the number of available rows in an array.
Ideally, we want to use a single array for high throughput. Thus, we pick the Pareto-optimal design with the least size while having MF no larger than the number of available rows in the target hardware.
When we must use multiple arrays, we pick the Pareto-optimal design with the least estimated EDP by considering all the instructions including the copy instructions.
\vspace{-1em}
\xadd{\begin{Example}
    In the example in Fig.~\ref{fig:form}, the red points correspond to the Pareto-optimal designs.
    If we want to use a single array with 8 available rows to compute the function, we pick design $D$ since among all the Pareto-optimal designs with MF no larger than 8, \textit{i.e.}, design $C$ and design $D$, design $D$ has a smaller size.
    Hence, it takes less time and energy to perform the operations in serial without the need of copy instructions.
    When we must use multiple arrays, we estimate the EDP of each of the three Pareto-optimal designs and pick the one with the least EDP.
\end{Example}}

Unlike EDP, which depends on hardware-specific features such as the row number of an array, size and MF do not depend on hardware, so we do not need to re-compile the function when using different IMC architectures, which is another advantage of using these two metrics as the optimization target.

%% file: 5_method.tex
\section{Method}\label{sec:method}
This section presents our logic compiler.
To obtain the Pareto-optimal designs on size and MF, we propose a compiler with multiple rounds of tightly coupled synthesis and scheduling.
To facilitate the illustration, we reorder the indexes of the operations in the netlist according to the execution sequence each time after scheduling to obtain the \emph{scheduled netlist}, where operation with index $i$ is executed at clock cycle $i$.
Since the MF of a scheduled netlist can be easily obtained using the method described in Section~\ref{sec:pre_comp}, a scheduled netlist contains all the information we need to measure the performance metrics of our interest, \textit{i.e.}, size and MF, of a design.
In what follows, we use the terms design and scheduled netlist interchangeably and use $\textit{Size}(G)$ and $\textit{MF}(G)$ to denote the size and MF of a scheduled netlist $G$, respectively.

\begin{figure}[!htbp]
\centering
\includegraphics[scale=0.4]{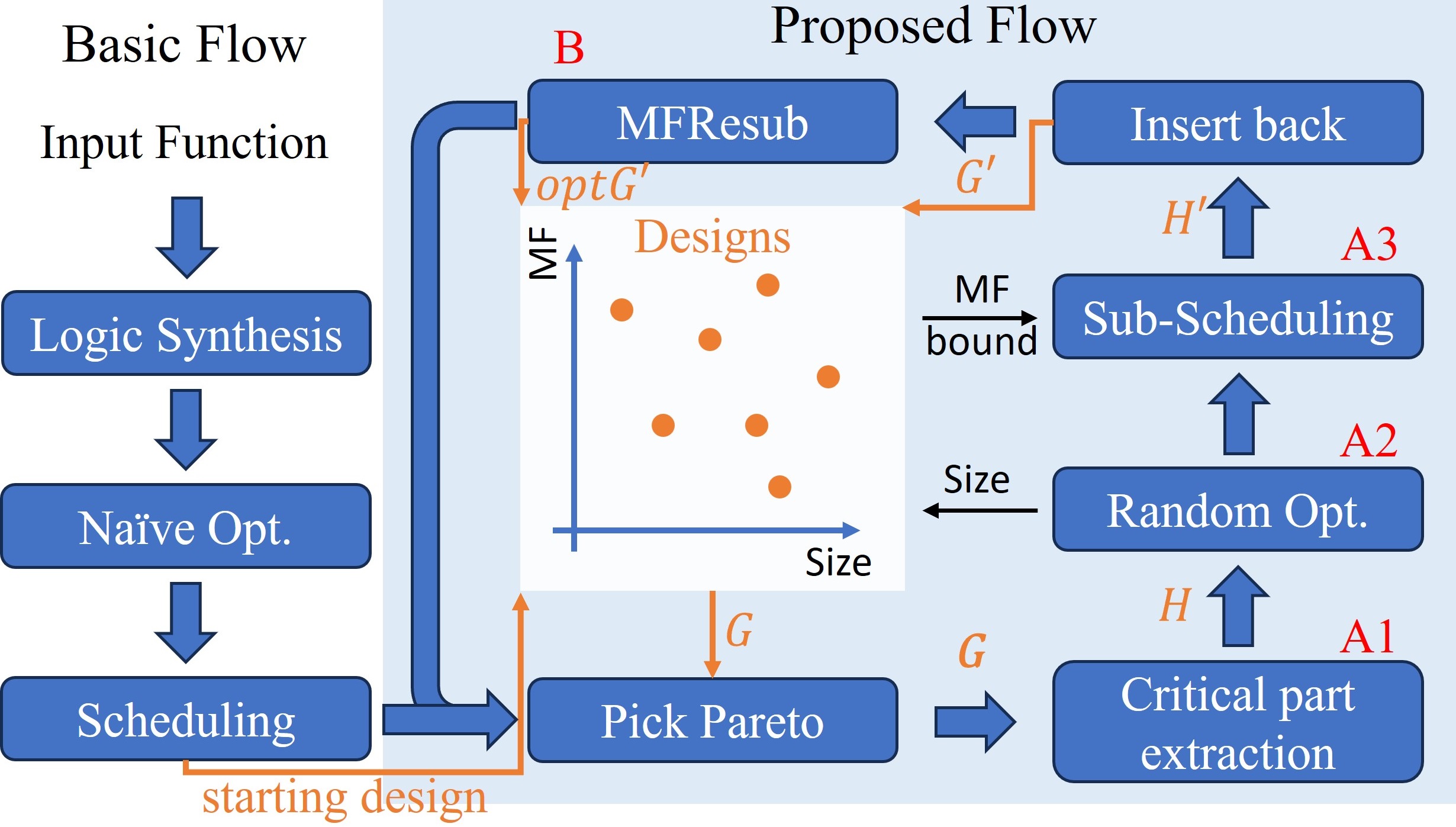}
\caption{The flow of our iterative logic compiler.}
\label{fig:propflow}
\end{figure}

Fig.~\ref{fig:propflow} shows the flow of our logic compiler.
First, the compiler executes a basic compiling flow to obtain a starting design.
Specifically, the input function is synthesized to an XMG netlist using Mockturtle~\cite{EPFLLIB}, and then optimized using the optimization commands in the tool to reduce the netlist size.
Then, the optimized netlist is scheduled with the state-of-the-art scheduler~\cite{PREV}, and the resulting design is added into the design space.
Now, instead of terminating the compiling flow as in most prior works~\cite{MIG,SIMPLER,STAR,LOGIC,PREV}, we try to find designs of better quality in terms of Pareto-optimality.
The \xdel{whole }process is shown in the light blue box in Fig.~\ref{fig:propflow}, and is run for $R$ rounds.
In each round, we randomly pick a baseline design $G$ from the Pareto-optimal designs of the current explored design space and try to improve it by modifying its netlist and re-scheduling it.
We pick from Pareto-optimal designs since they are the ones with the highest quality so far and thus, are promising candidates for optimization to advance the Pareto frontier.

Since the scheduler~\cite{PREV} and some netlist optimization commands are time-consuming for large netlists, we first exploit the scheduling information to extract a \emph{critical sub-netlist} $H$ from the netlist that is crucial for the quality improvement, and attempt to re-synthesize and re-schedule it instead of the whole netlist.
The optimized critical sub-netlist $H'$ will be inserted back into the baseline design $G$ to form a new design $G'$.
The details of the critical sub-netlist optimization will be introduced in Section~\ref{sec:method_crit}.
To further improve the quality of the design, we propose \emph{MFResub} in Section~\ref{sec:method_mfresub}, a resubstitution method oriented at MF reduction by taking the scheduling into account.
We apply \textit{MFResub} to the newly generated design $G'$ and then put the resulting design $\textit{opt}G'$ into the design space to update the Pareto frontier.
% Note that MFResub is applied to the whole netlist instead of the critical sub-netlist since it is relatively fast and we can find more optimization opportunities on the whole netlist.

\input{5_1_critical}
\input{5_2_MFResub}

%% file: 5_1_critical.tex
\subsection{Critical Sub-netlist Optimization}\label{sec:method_crit}
In order to efficiently improve an existing design by re-synthesis and re-scheduling, we propose to work on a critical sub-netlist instead of the whole netlist.
In the following, we will first explain how to identify and extract the critical sub-netlist in Section~\ref{sec:extract}.
Then, we will illustrate how to optimize and schedule it in Sections~\ref{sec:opt} and~\ref{sec:sched}, respectively.
After we get the optimized and re-scheduled sub-netlist, we will replace the old critical sub-netlist in the original design by it to get a new design for the input function.

\subsubsection{Critical Sub-netlist Extraction}\label{sec:extract}
\begin{figure}[!htbp]
\centering
\includegraphics[scale=0.8]{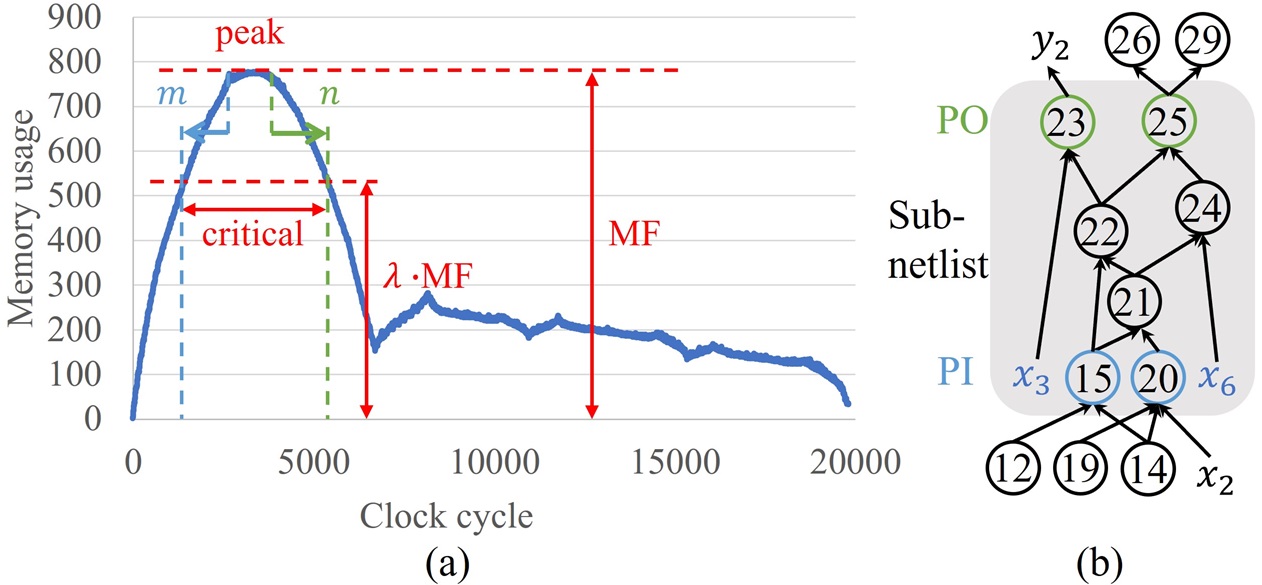}
\caption{(a) Memory usage and critical sub-netlist of \textit{log2} benchmark; (b) an example critical sub-netlist.}
\label{fig:extract}
\end{figure}
To identify the critical sub-netlist, we first define \emph{memory usage} at clock cycle $i$ as the number of occupied memory rows, \textit{i.e.}, the rows containing the operation results that cannot be overwritten, at the end of clock cycle $i$ after operation $i$ is scheduled.
% The memory usage at each clock cycle can be easily obtained from the scheduled netlist $G$ using the method described in Section~\ref{sec:pre_comp}.
In Fig.~\ref{fig:extract}(a), we plot the memory usage versus clock cycle of a netlist of benchmark \textit{log2}.
We can see that during the entire period, the memory usage starts at one, reaches the maximum value, which equals the MF of the scheduled netlist by definition, and ends at a value equal to the PO number.
% Note that in the general situation, the memory usage can reach the maximum value multiple times during the entire period.
We define \emph{peak period} as the period from clock cycle $m$ to clock\ cycle $n$, where $m$ and $n$ are the first and last clock cycles at which the memory usage reach MF, respectively\xadd{, as shown in Fig.~\ref{fig:extract}(a)}.
If we can decrease the maximum memory usage during the peak period, we decrease the MF of the whole netlist.
Since working with the operations scheduled during the peak period only is too restrictive, we expand the peak period to include some cycles before and after it.
Specifically, as illustrated in Fig.~\ref{fig:extract}(a), we decrease the value of $m$ and increase the value of $n$ until the memory usage at clock cycles $m$ and $n$ drops below $\lambda \cdot \text{MF}$, where $0<\lambda<1$ is called the \emph{critical parameter}.

The critical sub-netlist are constructed with the operations scheduled between the above updated clock cycles $m$ and $n$, \textit{i.e.}, those with indexes ranging from $m$ to $n$.
Besides, some nodes in the entire netlist $G$ should be added to serve as its PIs, and some nodes of the sub-netlist should be identified as its POs. Next, 
we show the construction process through an example\xdel{ in Fig~\ref{fig:extract}(b)}.
\vspace{-1em}
\begin{Example}
\xadd{In Fig~\ref{fig:extract}(b), we assume}\xdel{Assume} that $m=21$ and $n=25$.
We construct the sub-netlist $H$ in the gray box whose operations are those indexed from $21$ to $25$ in the scheduled netlist $G$.
The PIs of the entire netlist $G$ and the operations scheduled before clock cycle 21 that are the fan-ins of the operations in $H$, \textit{i.e.}, $x_3$, $x_6$, operation 15, and operation 20, are included in $H$ as its PIs.
The POs of $H$ are the operations of $H$ that have fan-outs outside the sub-netlist, \textit{i.e.}, operation 25, and operations that are POs of $G$, \textit{i.e.}, operation 23.
\end{Example}

The sub-netlist $H$ formed in this way can be extracted from $G$ as an individual netlist, and we can modify $H$ by optimization commands without affecting the function of $G$ as long as the functions of the POs of $H$ are unchanged. 
In this way, we can work on $H$ whose performance is crucial but with a much smaller size.

\subsubsection{Sub-netlist Optimization}\label{sec:opt}
We apply $K$ random optimization commands in Mockturtle~\cite{EPFLLIB} including resubstitution~\cite{WINDRESUB} and rewriting~\cite{REW1,REW2}, to the critical sub-netlist $H$, hoping to obtain a functionally equivalent netlist $H'$ with better quality\xdel{a smaller size}.
To further reduce the size, we additionally apply the powerful command sequence \textit{resyn2} in ABC~\cite{ABC} using the corresponding commands in Mockturtle~\cite{EPFLLIB} to $H'$.

\subsubsection{Sub-netlist Scheduling}\label{sec:sched}
We apply the scheduler from~\cite{PREV} to obtain the execution sequence of $H'$.
During the scheduling, the PIs of $H'$ that are the operations of $G$, \textit{e.g.}, operations 15 and 20 in Fig.~\ref{fig:extract}(b), are regarded as \emph{temporary inputs} of the sub-netlist defined in~\cite{PREV}, and they can be deleted from memory when they are no longer needed.
After scheduling, the indexes of the operations in $H'$ are also reordered according to the execution sequence.

To accelerate the process, we adopt an early-termination rule in the scheduling process.
Specifically, after we obtain the optimized sub-netlist $H'$, we can calculate the size of the netlist $G'$, which will be obtained later by substituting the critical sub-netlist $H$ of the original netlist $G$ with $H'$. Its size is
$$ \textit{Size}(G')=\textit{Size}(G)-\textit{Size}(H)+\textit{Size}(H').$$
Since we want to find Pareto-optimal designs, we can obtain an upper bound of MF so that $G'$ can be a new Pareto-optimal design.
Specifically, among all Pareto-optimal designs we have found so far, we identify the one, denoted as $D$, with size no more than $\textit{Size}(G')$ and closest to $\textit{Size}(G')$. Then, we require the MF of $G'$ to be less than $\textit{MF}(D)$.
    \vspace{-1em}
\xadd{\begin{Example}
For example, suppose that the designs we have explored so far are those shown in Fig.~\ref{fig:form} and that the size of $G'$ is 43.
Among all Pareto-optimal designs we have found so far, \textit{i.e.}, the red points, the one with size no more than $\textit{Size}(G')$ and closest to $\textit{Size}(G')$ is the design $D$ shown in Fig.~\ref{fig:form}. Since its MF is $8$, we require that the MF of $G'$ be less than 8 if it would become a Pareto-optimal design.
\end{Example}}
Since we will apply our MF-oriented resubstitution to $G'$ later to further improve its quality, we relax this bound a bit in scheduling stage by multiplying it with a \emph{bound relaxation parameter} $\beta>1$.

The above bound can be used in the scheduler in~\cite{PREV} to achieve the early termination.
% Actually, the scheduler in~\cite{PREV} first divides the input netlist into sub-netlists and then schedules each sub-netlist individually by a binary search flow that iteratively checks whether an MF upper bound is feasible for the sub-netlist.
% Thus, when scheduling each sub-netlist, we can input the above bound to the scheduler.
If the bound is not achievable, we terminate the scheduler immediately and begin another round of the compiler flow without executing the remaining steps in this round.

%% file: 5_2_MFResub.tex
\subsection{MF-Oriented Resubstitution}\label{sec:method_mfresub}
\begin{algorithm}\label{algo:mfresub}
\begin{small}
\caption{The flow of MF-oriented resubstitution.}
    \KwIn{$G$: a scheduled netlist;
    $N_{\textit{trial}}$: a limit on the number of trials.}
    \KwOut{$\textit{opt}G$: the optimized netlist.}
    $p\gets \textit{FirstPeak}(G)$\;\label{A2_p}
    $\mathcal{P}\gets \textit{OpInMem}(p)$\;\label{A2_peak}
    \ForEach{$j\in\mathcal{P}$}
    {\label{A2_eq_begin}
        \ForEach{$i\in\mathcal{P}$ \textbf{with} $i<j$}
        {\label{A2_eq_i}
            $s \gets \textit{CheckSub} (G,j,i)$\;\label{A2_eq_check}
            \lIf{$s$ is not \textit{NULL}}
            {\Return{$\textit{SubOp}(G,j,s)$}\label{A2_eq_ret}}
        }
    }\label{A2_eq_end}
    \ForEach{$j\in\mathcal{P}$}
    {\label{A2_fo_begin}
        \lIf{$\textit{isPO}(j)$ \textbf{or} $\textit{NoSingFO}(j,p)$}{\textbf{continue}}\label{A2_fo_filt}
        $f\gets \textit{GetSingFO}(j,p)$\;\label{A2_fo_get}
        $\textit{Divisors}\gets (\mathcal{P}\backslash\{j\})\cup \{k|p<k<f\}$\;\label{A2_fo_divisor}
        $\textit{Divisors}\gets \textit{PruneDiffPI}(\textit{Divisors},f) \cup \textit{GetPI}(f)$\;\label{A2_fo_prune}
        \If{$\textit{NumEnum}(|\textit{Divisors}|)\le N_{\textit{trial}}$}
        {\label{A2_fo_nenum}
            $\textit{DivTrip}\gets \textit{Enum}(\textit{Divisors})$\;\label{A2_fo_enum}
        }
        \lElse{$\textit{DivTrip}\gets \textit{Rand}(\textit{Divisors}, N_{\textit{trial}})$}\label{A2_fo_rand}
        \For{$(u,v,w)\in\textit{DivTrip}$}
        {\label{A2_fo_three_beg}
            $s \gets \textit{CheckSubOp} (G,f,(u,v,w)))$\;\label{A2_fo_check}
           \lIf{$s$ is not \textit{NULL}}
            {\Return{$\textit{SubOp}(G,f,s)$}\label{A2_fo_ret}}
        }\label{A2_fo_three_end}
    }\label{A2_fo_end}
    \Return{$G$}\;\label{A2_noresub}
\end{small}
\end{algorithm}
\begin{figure}[!htbp]
\centering
\includegraphics[scale=0.35]{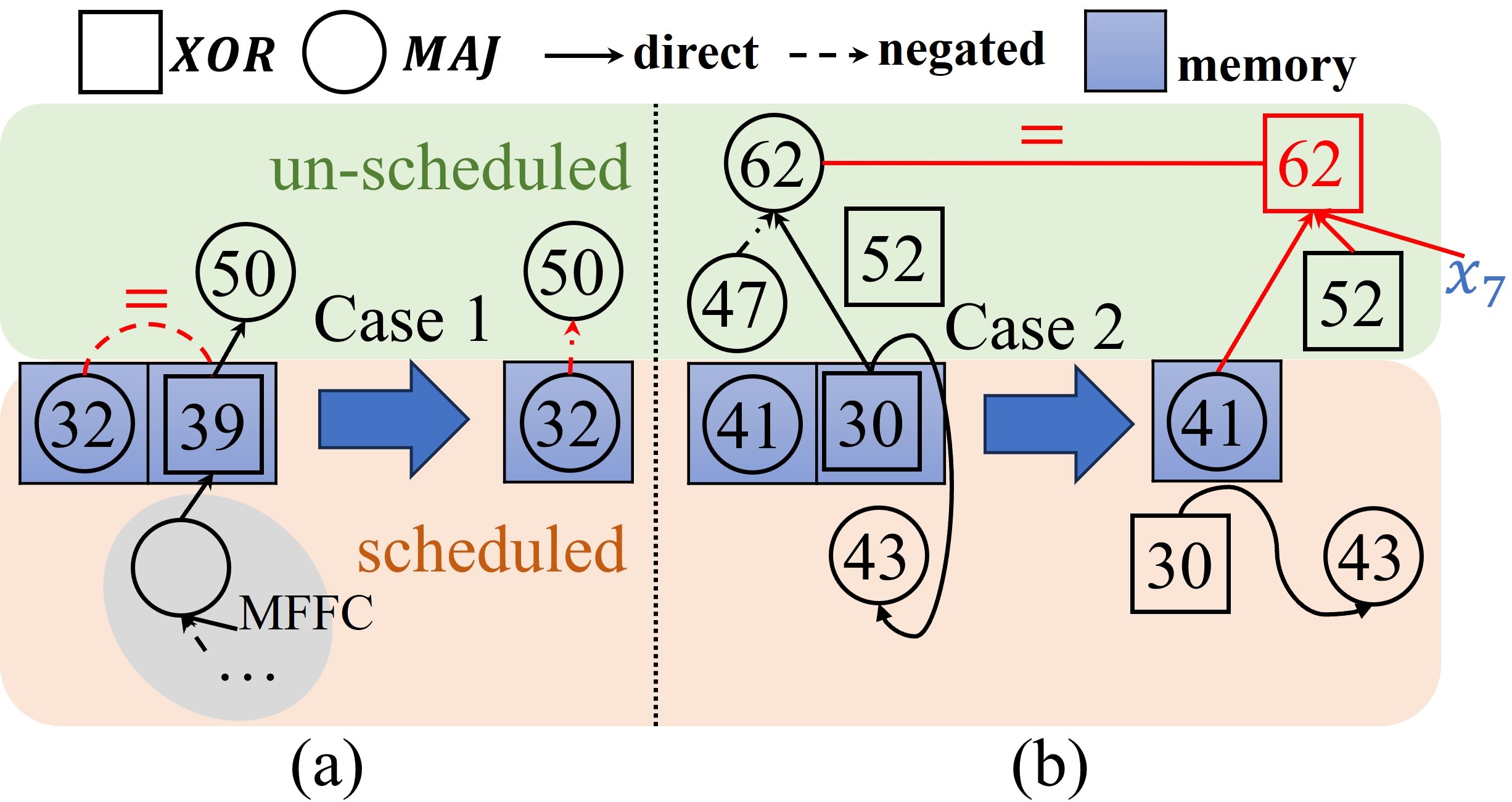}
\caption{The two cases of MF-oriented resubstitution\xadd{: (a) case 1; (b) case 2}.}
\label{fig:mfresub}
\end{figure}
To further improve the quality of a design, we propose $\textit{MFResub}$, a resubstitution method oriented at MF reduction by taking the scheduling into account.
The pseudocode is shown in Algorithm~\ref{algo:mfresub}, and we will explain it using the two examples in Fig.~\ref{fig:mfresub}.
The inputs are a scheduled netlist $G$ and a limit on the number of trials $N_{\textit{trial}}$.
The output is the optimized netlist $\textit{opt}G$.
Line~\ref{A2_p} obtains $p$, the first clock cycle at which the memory usage reaches the maximum value, \textit{i.e.}, the MF of the netlist.
Line~\ref{A2_peak} obtains the operations whose results are stored in the memory at clock cycle $p$ after operation $p$ is scheduled and stores them in a set $\mathcal{P}$.
By definition, the size of $\mathcal{P}$ equals the MF of the netlist.
If we can delete the result of an operation in $\mathcal{P}$ from memory at clock cycle $p$ by resubstitution, we may decrease the MF of the whole netlist.
In the examples in Fig.~\ref{fig:mfresub}, $p=45$, and the set $\mathcal{P}$ is indicated using the blue boxes.
We attempt to find the following two cases of resubstitution.

\subsubsection{Case 1}
Lines~\ref{A2_eq_begin}--\ref{A2_eq_end} correspond to case one as shown in\xdel{ the left of} Fig.~\ref{fig:mfresub}\xadd{(a)}.
In case one, two operations stored in memory \xdel{have the same global functions in terms of the PIs}are functionally equivalent, so we can remove one of them from the netlist, thus deleting its result from the memory.
Line~\ref{A2_eq_check} checks whether an operation $j$ in $\mathcal{P}$ is functionally equivalent to another (possibly negated) operation $i$ in $\mathcal{P}$ with smaller index with a function $\textit{CheckSub}$, which uses the checking process of the simulation-guided resubstitution~\cite{SIMRESUB} described in Section~\ref{sec:back_resub}.
If so, $\textit{CheckSub}$ returns $s$ as $i$ or its negation according to the equivalence result.
Otherwise, it returns $s$ as \textit{NULL}.
% For example, operation 39 is functionally equivalent to the negation of operation 32, which is returned as $s$.
If $s$ is not \textit{NULL}, function $\textit{SubOp}$ at Line~\ref{A2_eq_ret} substitutes the operation $j$ with $s$ and removes all operations in the MFFC of $j$.
With $j$ removed from the netlist, its result can also be deleted from memory, which may lead to a reduction in MF.
\vspace{-1em}
\xadd{\begin{Example}
In the example in Fig.~\ref{fig:mfresub}\xadd{(a)}, among the operations whose results are stored in memory, operation 39 is functionally equivalent to the negation of operation 32.
Hence, we substitute operation 39 with the negation of operation 32 when it serves as a fan-in of node 50.
After the substitution, we remove node 39 from the netlist together with its MFFC, leading to a reduction in netlist size and the current memory usage, which may lead to a reduction in MF.
\end{Example}}

\subsubsection{Case 2}
Lines~\ref{A2_fo_begin}--\ref{A2_fo_end} correspond to case two as shown in\xdel{ the right of} Fig.~\ref{fig:mfresub}\xadd{(b)}.
In case two, we cannot remove an operation from the netlist, but we can resubtitute its fan-out to make its result not needed by any un-scheduled operations.
Thus, we can delete its result from the memory.
We iterate over all operations $j\in\mathcal{P}$ and focus on the operation that is not a PO and has only one un-scheduled fan-out at clock cycle $p$\xdel{, \textit{e.g.}, operation 30}.
Line~\ref{A2_fo_filt} prunes the unqualified operations. For a qualified operation, Line~\ref{A2_fo_get} obtains its single unscheduled fan-out $f$\xdel{, \textit{e.g.}, operation 62}.
If we can substitute $f$ with a new operation that does not take $j$ as its fan-in, the result of $j$ is no longer needed and can be deleted from memory at clock cycle $p$.
The alternative fan-ins of this new operation are picked from a set $\textit{Divisors}$ constructed in Lines~\ref{A2_fo_divisor}--\ref{A2_fo_prune}.
Since we do not want to change the execution sequence of the netlist, Line~\ref{A2_fo_divisor} initially constructs the set to include all the other operations whose results are in memory at clock cycle $p$, \textit{i.e.}, $\mathcal{P}\backslash\{j\}$, and the operations to be scheduled from clock cycle $(p+1)$ to clock cycle $(f-1)$.
Then, Line~\ref{A2_fo_prune} refines the set based on PI dependency.
We say that an operation depends on a PI if there is a directed path from the PI to the operation.
The function of an operation is entirely determined by the PIs on which it depends.
If an operation $d\in \textit{Divisors}$ depends on a PI $u$, but the operation $f$ does not depend on $u$, it is unlikely that we can find a functionally equivalent new operation with $d$ as its fan-in to substitute $f$.
Hence, Line~\ref{A2_fo_prune} prunes these divisors by the function $\textit{PruneDiffPI}$. Besides, it also puts the PIs on which $f$ depends and the constant signal into $\textit{Divisors}$.

Next, since our netlist is XMG where each operation has three inputs, we need to check whether each triple $(u,v,w)$ where $u,v,w\in \textit{Divisors}$ can be used as the fan-ins of the new operation to substitute $f$.
We put all the triples we will try into a set $\textit{DivTrip}$ and bound its size with a parameter $N_{\textit{trial}}$.
If the number of different triples in $\textit{Divisors}$ is no larger than $N_{\textit{trial}}$, we enumerate them in Line~\ref{A2_fo_enum}.
Otherwise, we randomly form $N_{\textit{trial}}$ triples in Line~\ref{A2_fo_rand}.
For each triple $(u,v,w)\in\textit{DivTrip}$, we use the function $\textit{CheckSubOp}$ to check whether a new operation $s$ of the supported type, \textit{i.e.}, MAJ or XOR, can be constructed using (possibly negated) $u,v,w$ as its fan-ins that is functionally equivalent to the operation $f$ in Line~\ref{A2_fo_check}.
The function $\textit{CheckSubOp}$ also uses the checking process in~\cite{SIMRESUB} to check for the equivalence.
% For example, operation 62 can be substituted with an XOR operation taking operation 41 stored in memory, operation 52 to be scheduled before operation 62, and PI $x_7$ as fan-ins.
If such an operation $s$ can be constructed, \textit{i.e.}, $s$ returned by $\textit{CheckSubOp}$ is not \textit{NULL}, we substitute the operation $f$ with $s$ in Line~\ref{A2_fo_ret}. Thus, we can remove $j$ from~$\mathcal{P}$.
\vspace{-1em}
\xadd{\begin{Example}
In the example in Fig.~\ref{fig:mfresub}\xadd{(b)}, operation 62 is functionally equivalent to a newly-constructed XOR operation taking operation 41 stored in memory, operation 52 to be scheduled before operation 62, and PI $x_7$ as fan-ins.
So we substitute operation 62 with the newly constructed operation (also indexed 62).
Since the original operation 62 is the only un-scheduled fan-out of operation 30, operation 30 is no longer needed for the future scheduling after the substitution.
Hence, we remove its result from memory, leading to a reduction in memory usage.
Note that the MFFC of the original operation 62 can be removed from the netlist\xadd{, leading to a reduction in netlist size}.
\end{Example}}

Note that during the process, the relative order of the indexes of the operations left in the netlist is not changed, so we can simply re-assign the indexes to skip those of the removed operations without re-scheduling the netlist.
In our implementation, if a resubstitution is found, the algorithm immediately returns the modified netlist (see Lines~\ref{A2_eq_end} and~\ref{A2_fo_ret}), and the entire algorithm is applied on the the modified netlist again. This process repeats until no resubstitution is found. In this case, the algorithm returns the input design in Line~\ref{A2_noresub}.

%% file: 6_results.tex
\section{Experimental Results}\label{sec:result}
This section shows the experimental results. We implement our logic compiler in C++ and perform experiments on a computer with a 24-core 2.4GHz Intel 4214R processor and 64GB RAM.
Our target hardware is the one proposed in~\cite{XMG-GPPIC}, and we use the \xdel{provided }data provided in~\cite{XMG-GPPIC} to estimate the EDPs of the designs.
Note that our compiler can be easily adapted to support other hardware by modifying the supported operations in \textit{MFResub} and using other logic synthesis tools such as ABC~\cite{ABC} instead of Mockturtle~\cite{EPFLLIB} to support the netlist type, \textit{e.g.}, NOR2 graph of the MAGIC architecture~\cite{MAGIC}, without changing our algorithms. 
The benchmarks used in our experiments are from EPFL benchmark suite~\cite{EPFLBENCH}, whose information is shown in Table~\ref{tab:bench}. \#Nodes are the number of nodes in the And-inverter graphs (AIGs) provided in the benchmark suite, and constant POs are not counted in \#POs.
\begin{table}[H]
  \centering
  \tabcolsep=2pt
  \caption{Benchmarks used in the experiment.}
    \begin{tabular}{cccccccc}
    \hline
      Name  & \#Nodes & \#PI  & \#PO  &  Name   & \#Nodes & \#PI  & \#PO \\
      \hline
    router & 257   & 60    & 3     & sin   & 5416  & 24    & 25 \\
    int2float & 260   & 11    & 7     & sqrt  & 24618 & 128   & 64 \\
    cavlc & 693   & 10    & 11    & multiplier & 27062 & 128   & 128 \\
    priority & 978   & 128   & 8     & log2  & 32060 & 32    & 32 \\
    max   & 2865  & 512   & 130   & div   & 57247 & 128   & 128 \\
    \hline
    \end{tabular}%
  \label{tab:bench}%
\end{table}

\begin{table*}
  \centering
  \caption{EDPs of the designs obtained by various compilers.}
\scriptsize
\begin{tabular}{ccccccccccccccccc}
    \toprule
    \multicolumn{1}{c}{\multirow{2}{*}{Benchmark}} & \multicolumn{3}{c}{GPPIC~\cite{XMG-GPPIC}} & \multicolumn{3}{c}{GPPIC+} & \multicolumn{3}{c}{Best BC (\cite{MIG,SIMPLER,STAR,PREV})} & \multicolumn{3}{c}{CTP} & \multicolumn{4}{c}{Ours} \\
\cmidrule(r){2-4} \cmidrule(r){5-7} \cmidrule(r){8-10} \cmidrule(r){11-13} \cmidrule(r){14-17}        & Size  & MF    & EDP   & Size  & MF    & EDP   & Size  & MF    & EDP   & Size  & MF    & EDP   & Size  & MF    & EDP   & Time(h) \\
    \midrule
    max   & 1935  & 336   & 0.93  & 1935  & 263   & 0.77  & 1824  & 264   & 0.72  & 2003  & 264   & 0.78  & 1820  & 263   & \textbf{0.71 } & 0.44 \\
    \midrule
    sin   & 3619  & 344   & 1.01  & 3619  & 262   & 0.87  & 3482  & 264   & 0.81  & 3540  & 252   & 0.80  & 3524  & 222   & \textbf{0.63 } & 1.32 \\
    \midrule
    sqrt  & 9699  & 194   & 11.16  & 9699  & 191   & 9.52  & 9240  & 275   & 15.34  & 9997  & 191   & 9.72  & 9166  & 188   & \textbf{8.12 } & 2.16 \\
    \midrule
    multiplier & 14251 & 1414  & 65.78  & 14251 & 261   & 33.68  & 14176 & 247   & \textbf{25.88 } & 14303 & 262   & 30.63  & 14303 & 253   & 26.31  & 4.71 \\
    \midrule
    log2  & 21276 & 1283  & 81.71  & 21276 & 763   & 46.64  & 19760 & 789   & 43.35  & 19821 & 771   & 41.81  & 19899 & 722   & \textbf{41.16 } & 3.92 \\
    \midrule
    div   & 28379 & 491   & 128.93  & 28379 & 272   & 92.43  & 12536 & 268   & 27.13  & 15272 & 377   & 34.27  & 12533 & 248   & \textbf{21.01 } & 4.79 \\
    \midrule
    GEOMEAN & 9144  & 521   & 13.93  & 9144  & 299   & 9.87  & 7685  & 316   & 8.06  & 8213  & 315   & 8.02  & 7708  & 281   & 6.61  & 2.19  \\
    \bottomrule
    \end{tabular}%
  \label{tab:edp}%
\end{table*}%
We set the number of rounds $R=20$, the critical parameter $\lambda = 0.6$, the bound relaxation parameter $\beta = 1.1$, the number of optimization commands $K=10$, and the limit on the number of trials $N_{\textit{trial}} = 500000$.
We compare our compiler to the following compilers.
\begin{itemize}
    \item
    GPPIC~\cite{XMG-GPPIC} and
    GPPIC+: GPPIC+ is an improved version of GPPIC. Since the scheduler used in our work is better than that used in GPPIC, we re-schedule the netlist obtained by GPPIC with the scheduler we use, \textit{i.e.},~\cite{PREV}, and call the combined approach GPPIC+.
    \item 
    Logic compilers that are based on basic compiling flows (BCs), \textit{i.e.},~\cite{MIG,SIMPLER,STAR,PREV}:
    These compilers target at different hardware architectures that support different types of netlists.
    In our comparison, we modify them to support XMG without changing their flows and algorithms. 
    % , so they use optimization commands provided in different tools to reduce the numbers of operations in the corresponding netlists in the synthesis step.
    % In our comparison, we modify them to support our target hardware from~\cite{XMG-GPPIC}, which handles XMG netlists.
    % Specifically, we replace their synthesis procedure by a strong synthesis procedure to minimize the XMG size as much as possible, \textit{i.e.}, repeatedly applying the Mockturtle version of \textit{resyn2} command as described in Section~\ref{sec:opt} until the size of the netlist cannot be decreased any more.
    % Their schedulers are still those used in their original versions.
    \item 
    A counterpart of our compiler (CTP):
    To show the strength of our flow featuring coupled synthesis and scheduling, we develop a counterpart of our compiler, \textit{i.e.}, CTP.
    CTP also iteratively optimizes the Pareto-optimal designs, but it performs the netlist optimization step in Section~\ref{sec:opt} on the whole netlist and schedules the resulting netlist.
    % However, instead of going through the coupled flow of synthesis and scheduling by optimizing critical sub-netlist and performing MF-oriented resubstitution, CTP performs the netlist optimization step in Section~\ref{sec:opt} on the whole netlist and schedules the resulting netlist.
    To make the comparison fair, we constrain the run time of CTP to be slightly longer than that of our compiler for each benchmark.
    % Since it takes longer time to modify and schedule the entire netlist, CTP can only run for less than $R$ rounds.
\end{itemize}

We set the number of rows in a memory array as 256. Under this size limit, four benchmarks can be computed within a single array.
That is, the sum of the number of PIs and MF is no larger than 256.
For these benchmarks, we compare the abilities of the compilers to obtain Pareto-optimal designs in Section~\ref{sec:result_compiler}.
For the other benchmarks, we compare the EDPs of the designs obtained by different compilers in Section~\ref{sec:result_EDP}.
In Section~\ref{sec:result_mfresub}, we further study the performance of our MF-oriented resubstitution.
\subsection{Small Benchmarks Using A Single Array}\label{sec:result_compiler}
\begin{figure}[!htbp]
\centering
\includegraphics[scale=0.72]{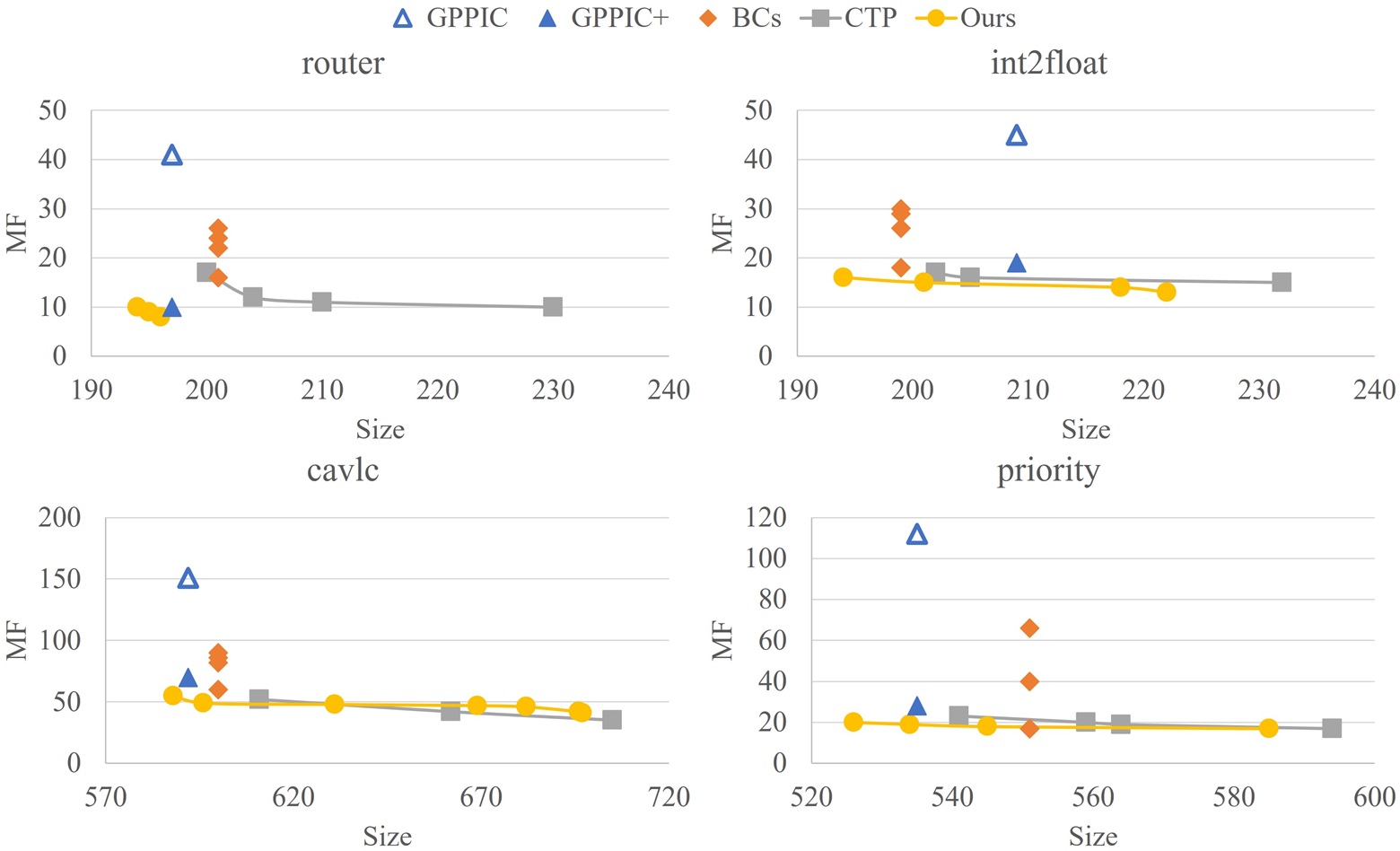}
\caption{Designs obtained by various compilers for small benchmarks.}
\label{fig:reslut_pareto}
\end{figure}
Fig.~\ref{fig:reslut_pareto} shows the designs obtained by various logic compilers on MF-size plane for the small benchmarks.
% For each benchmark, the hollow and solid triangles are the designs obtained by GPPIC and GPPIC+, respectively. They share the same netlist optimized by GPPIC~\cite{XMG-GPPIC}, but scheduled with different schedulers, \textit{i.e.}, the original one in~\cite{XMG-GPPIC} and the one proposed in~\cite{PREV}, so they have the same size and the design scheduled by~\cite{PREV} has less MF.
% Similarly, the four orange diamonds are the designs obtained by the four BCs. They share the same netlist optimized by iteratively applying \textit{resyn2} commands, but scheduled with different schedulers proposed in~\cite{MIG,SIMPLER,STAR,PREV}, so they also have the same size.
% Instead of only obtaining one design, our compiler and its counterpart CTP can find several designs shown as yellow circles and gray rectangles, respectively.
We have shown in Section~\ref{sec:mot} that it is important to find high quality designs in terms of Pareto-optimality on size and MF.
Hence, a good compiler is the one that can find designs lying in the bottom left corner.
From the figure, we can see that in general, our compiler outperforms the others.
% in finding Pareto-optimal designs.
\subsection{Large Benchmarks Using Multiple Arrays}\label{sec:result_EDP}
For a large benchmark, we must use multiple arrays to compute the function.
Cross-array data copying instructions are inserted into the instruction sequence when needed.
We use the method in~\cite{XMG-GPPIC} to estimate the EDP ($10^{-15}Js$) of the designs by accumulating the energy and delay of each instruction, and show the result in Table~\ref{tab:edp}.
For simplicity, for each benchmark, we present the best BC design, which is the design with the least EDP obtained by the four BCs.
% We have shown in Section~\ref{sec:mot} that the Pareto-optimal designs tend to have less EDP.
For our compiler and CTP, we estimate the EDPs of the Pareto-optimal designs obtained and pick the one with the least EDP.
% Note that our method is agnostic of the number of rows in an array, so the Pareto-optimal designs obtained by our compiler can be applied to the hardware with other array sizes without re-compiling.

The compiler giving the smallest EDP for each benchmark is marked in bold.
We can see that our compiler outperforms the others for all benchmarks except one.
Our compiler reduces the geometric mean of EDPs by 52.6\%, 33.0\%, 18.0\%, and 17.6\% compared to GPPIC~\cite{XMG-GPPIC}, GPPIC+, the best result of the compilers with basic flow, \textit{i.e.},~\cite{MIG,SIMPLER,STAR,PREV}, and the counterpart of our compiler.
For the run time, our compiler can finish within 5 hours for all the benchmarks, while GPPIC requires more than 8 hours to compile large benchmarks on a computer with a similar performance.

% Now, we perform a more detailed analysis on the experimental results.
As shown in Table~\ref{tab:edp}, GPPIC~\cite{XMG-GPPIC} and its improved version cannot achieve good performance, since their optimization command sequence only contains 10 commands.
This shows the importance of running the iterative improvement flow in the light blue box of Fig.~\ref{fig:propflow} for $R$ rounds in our compiler.
For the BCs, the netlist size is effectively reduced:
As we can see from the table, the geometric mean of the sizes of the netlists obtained by the BCs is the smallest among all the methods.
However, a design with \xdel{a }smaller size does not necessarily has \xdel{a }less EDP.
For example, for \textit{sin} benchmark, our design has larger size compared to the best BC design, but it has 23\% less EDP.
This result further confirms the advantage of an iterative improvement flow, where a scheduled netlist is re-synthesized to improve its performance, over those one-pass flows.
Our compiler also outperforms its counterpart which optimizes and schedules the whole netlist.
This shows the strength of tight coupling of synthesis and scheduling in the iterative flow through our proposed techniques such as critical sub-netlist extraction and MF-oriented resubstitution, which enhance\xdel{s} the efficiency of the flow.

\subsection{Performance of $\textit{MFResub}$}\label{sec:result_mfresub}
\begin{figure}[!htbp]
\centering
\includegraphics[scale=0.5]{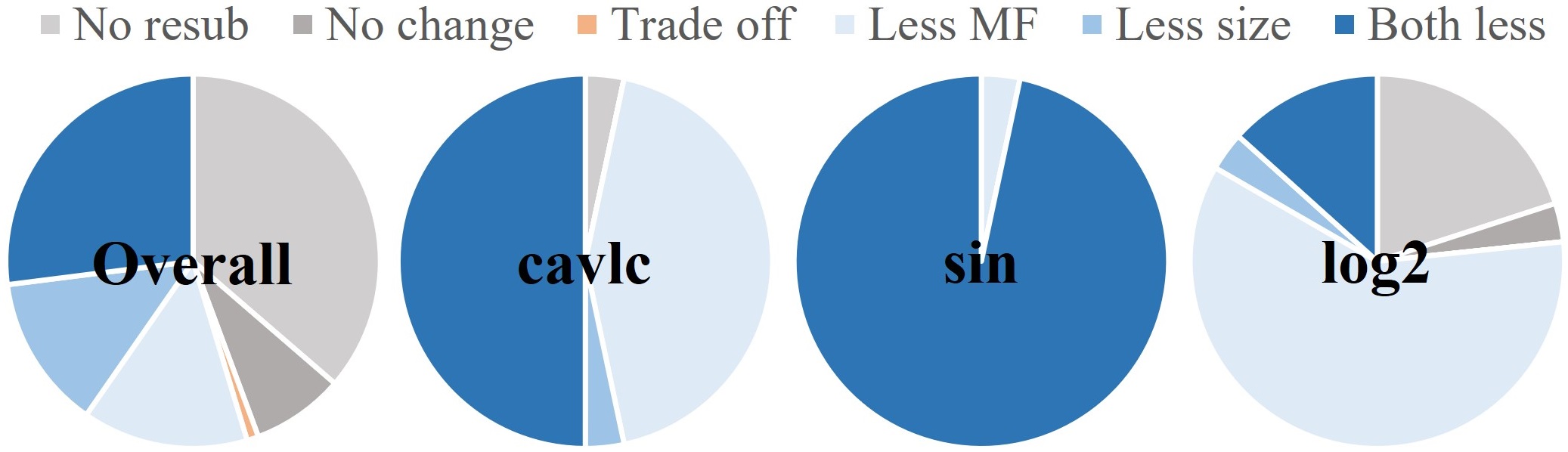}
\caption{The performance of $\textit{MFResub}$.}
\label{fig:reslut_mfresub}
\end{figure}
In this section, we analyze the performance of our MF-oriented resubstitution $\textit{MFResub}$.
We generate 30 netlists for each benchmark with random optimization commands in Mockturtle~\cite{EPFLLIB}, and apply $\textit{MFResub}$ to them.
We categorize the outcome into the following six categories and show it as pie charts in Fig.~\ref{fig:reslut_mfresub}. 
1) No resub: no resubstitution is made in the process.
2) No change: resubstitution is made on the netlist but it does not change its size or MF.
3) Trade off: in a very rare case, the resulting netlist has smaller size but slightly larger MF.
4) Less MF: the resulting netlist has the same size and less MF.
5) Less size: the resulting netlist has the same MF and smaller size.
6) Both less: the resulting netlist has smaller size and less MF.
Since our target is to find Pareto-optimal designs on size and MF, the latter three categories shown in blue colors all can be regarded as quality improvement.

We can see\xadd{ from Fig.~\ref{fig:reslut_mfresub}} that our $\textit{MFResub}$ can reduce both the size and MF for over a quarter of the $30\times10$ netlists and improve the quality for more than half netlists, but fail to find any resubstitution for a third of the netlists.
The performance differs greatly on different benchmarks.
The algorithm can perform well on netlists of various sizes.
For example, $\textit{MFResub}$ can improve the quality of most netlists of \textit{cavlc}, \textit{sin}, and \textit{log2} benchmarks as shown in the figure.
However, for some other benchmarks such as \textit{sqrt}, it can hardly make any progress.
Our current $\textit{MFResub}$ only performs MF-oriented resubstitution in two cases as described in Section~\ref{sec:method_mfresub}.
In the future, we will identify more cases where resubstitution can be made to reduce size and MF of the scheduled netlist.

%% file: 7_conclusion.tex
\section{Conclusion}\label{sec:conclusion}
% In this work, we build a high-quality iterative logic compiler for SIMD IMC with tight coupling of synthesis and scheduling, which is based on improving the critical sub-netlist and performing MF-oriented resubstitution.
% The experimental results show that our compiler can obtain designs of better quality and lower EDPs than state-of-the-art methods.
In this work, we build a high-quality compiler for SIMD IMC.
We identify two metrics, \textit{i.e.}, size and MF, of a scheduled netlist that are crucial for the end-to-end performance after applying the design on hardware.
We reveal that netlists with smaller sizes do not necessarily have less MF.
Hence, the target of our compiler is to find high-quality designs in terms of Pareto-optimality on size and MF.
We propose an iterative compiler with tight coupling of synthesis and scheduling, which is based on improving the critical sub-netlist identified by our algorithm using scheduling result and performing MF-oriented resubstitution that can reduce size and MF of a scheduled netlist without the need of re-scheduling.
The experimental results show that our compiler can obtain designs of better quality and the geometric mean of the EDPs of the designs is reduced by 18.0\% compared to the best state-of-the-art method.
\xadd{In the future, we will identify more cases where resubstitution can be made to reduce size and MF of the scheduled netlist in our MF-oriented resubstitution method to further improve the performance of our compiler.}

%% file: main.bbl
\begin{thebibliography}{10}
\providecommand{\url}[1]{#1}
\csname url@samestyle\endcsname
\providecommand{\newblock}{\relax}
\providecommand{\bibinfo}[2]{#2}
\providecommand{\BIBentrySTDinterwordspacing}{\spaceskip=0pt\relax}
\providecommand{\BIBentryALTinterwordstretchfactor}{4}
\providecommand{\BIBentryALTinterwordspacing}{\spaceskip=\fontdimen2\font plus
\BIBentryALTinterwordstretchfactor\fontdimen3\font minus \fontdimen4\font\relax}
\providecommand{\BIBforeignlanguage}[2]{{%
\expandafter\ifx\csname l@#1\endcsname\relax
\typeout{** WARNING: IEEEtran.bst: No hyphenation pattern has been}%
\typeout{** loaded for the language `#1'. Using the pattern for}%
\typeout{** the default language instead.}%
\else
\language=\csname l@#1\endcsname
\fi
#2}}
\providecommand{\BIBdecl}{\relax}
\BIBdecl

\bibitem{WALL}
A.~Pedram \emph{et~al.}, ``Dark memory and accelerator-rich system optimization in the dark silicon era,'' \emph{IEEE Design \& Test}, vol.~34, no.~2, pp. 39--50, 2017.

\bibitem{MAGIC}
S.~Kvatinsky \emph{et~al.}, ``{MAGIC}--memristor-aided logic,'' \emph{IEEE Trans. Circuits Syst. II, Exp. Briefs}, vol.~61, no.~11, pp. 895--899, 2014.

\bibitem{SIMDRAM}
N.~Hajinazar \emph{et~al.}, ``{SIMDRAM}: A framework for bit-serial {SIMD} processing using {DRAM},'' in \emph{ASPLOS}, 2021, pp. 329--345.

\bibitem{PLIM}
P.-E. Gaillardon \emph{et~al.}, ``The programmable logic-in-memory ({PLiM}) computer,'' in \emph{DATE}, 2016, p. 427–432.

\bibitem{XMG-GPPIC}
C.~Nie \emph{et~al.}, ``{XMG-GPPIC}: Efficient and robust general-purpose processing-in-cache with {XOR-Majority-Graph},'' in \emph{GLSVLSI}, 2023, pp. 183--187.

\bibitem{28NM}
J.~Wang \emph{et~al.}, ``A 28-nm compute {SRAM} with bit-serial logic/arithmetic operations for programmable in-memory vector computing,'' \emph{IEEE J SOLID-ST CIRC}, vol.~55, no.~1, pp. 76--86, 2020.

\bibitem{MIG}
M.~Soeken \emph{et~al.}, ``An {MIG}-based compiler for programmable logic-in-memory architectures,'' in \emph{DAC}, 2016, pp. 1--6.

\bibitem{SIMPLER}
R.~Ben-Hur \emph{et~al.}, ``{SIMPLER MAGIC}: Synthesis and mapping of in-memory logic executed in a single row to improve throughput,'' \emph{TCAD}, vol.~39, no.~10, pp. 2434--2447, 2020.

\bibitem{STAR}
F.~Wang \emph{et~al.}, ``{STAR}: Synthesis of stateful logic in {RRAM} targeting high area utilization,'' \emph{TCAD}, vol.~40, no.~5, pp. 864--877, 2021.

\bibitem{PREV}
X.~Qian \emph{et~al.}, ``An efficient logic operation scheduler for minimizing memory footprint of in-memory {SIMD} computation,'' in \emph{DATE}, 2024.

\bibitem{LOGIC}
M.~R.~H. Rashed \emph{et~al.}, ``Logic synthesis for digital in-memory computing,'' in \emph{ICCAD}, 2022, pp. 1--9.

\bibitem{SYN}
S.~Shirinzadeh \emph{et~al.}, ``Logic synthesis for in-memory computing using resistive memories,'' in \emph{ISVLSI}, 2018, pp. 375--380.

\bibitem{CHAL}
N.~Talati \emph{et~al.}, ``Practical challenges in delivering the promises of real processing-in-memory machines,'' in \emph{DATE}, 2018, pp. 1628--1633.

\bibitem{REW1}
H.~Riener \emph{et~al.}, ``On-the-fly and {DAG}-aware: Rewriting boolean networks with exact synthesis,'' in \emph{DATE}, 2019, pp. 1649--1654.

\bibitem{ABC}
B.~Brayton \emph{et~al.}, ``{ABC}: An academic industrial-strength verification tool,'' in \emph{Proceedings of the 22nd International Conference on Computer Aided Verification}, 2010, p. 24–40.

\bibitem{SIMRESUB}
S.-Y. Lee \emph{et~al.}, ``A simulation-guided paradigm for logic synthesis and verification,'' \emph{TCAD}, vol.~41, no.~8, pp. 2573--2586, 2022.

\bibitem{NVSIM}
X.~Dong \emph{et~al.}, ``{NVSim}: A circuit-level performance, energy, and area model for emerging nonvolatile memory,'' \emph{{TCAD}}, vol.~31, no.~7, pp. 994--1007, 2012.

\bibitem{WINDRESUB}
A.~Mishchenko \emph{et~al.}, ``Scalable logic synthesis using a simple circuit structure,'' in \emph{IWLS}, 2006.

\bibitem{EPFLLIB}
M.~Soeken \emph{et~al.}, ``The {EPFL} logic synthesis libraries,'' 2022, arXiv:1805.05121v3.

\bibitem{REW2}
H.~Riener \emph{et~al.}, ``Boolean rewriting strikes back: Reconvergence-driven windowing meets resynthesis,'' in \emph{ASP-DAC}, 2022, pp. 395--402.

\bibitem{EPFLBENCH}
L.~Amarù \emph{et~al.}, ``The {EPFL} combinational benchmark suite,'' in \emph{IWLS}, 2015.

\end{thebibliography}
